\documentclass[prc,twocolumn,nofootinbib]{revtex4-2}

\usepackage{color}
\usepackage{graphicx}% Include figure files
\usepackage{dcolumn}% Align table columns on decimal point
\newcolumntype{d}[1]{D{.}{.}{#1}}
\usepackage{bm}% bold math
\usepackage{braket}
\usepackage{amsmath}
\usepackage{here}
\usepackage{color}
\usepackage{array}
\usepackage{enumerate}
\usepackage{amssymb}
\usepackage{bbm}
\usepackage{dsfont}
\usepackage[scr=rsfs]{mathalpha}
\usepackage{longtable}

\def\be{\begin{equation}} 
\def\ee{\end{equation}}

\renewcommand{\vec}[1]{\mbox{\boldmath $#1$}}
\usepackage{xcolor}
\graphicspath{{figs/}}
\usepackage[bookmarks=true,colorlinks,%
citecolor=blue,linkcolor=blue,anchorcolor=blue,filecolor=blue,urlcolor=blue,%
           ]{hyperref}

\usepackage[normalem]{ulem}  % \sout{old text} for strikeout

\renewcommand\sout{\bgroup\color[rgb]{1,0.75,0.8} \ULdepth=-.5ex \ULset}
\begin{document}

\title{
Non-equilibrium Green's function approach to low-energy fission 
dynamics: fluctuations in fission reactions}

\author{K. Uzawa}
\affiliation{%
 Department of Physics, Kyoto University, Kyoto 606-8502, Japan 
}%

\author{K. Hagino}
\affiliation{%
 Department of Physics, Kyoto University, Kyoto 606-8502, Japan 
}%

\begin{abstract}
We present a microscopic modeling for a decay of a heavy 
compound nucleus, starting from a nucleonic degree of 
freedom. 
To this end, we develop an approach based on a non-equilibrium Green's function, which is combined 
with a configuration interaction (CI) approach based on a constrained density-functional theory (DFT). 
We apply this approach to a barrier-top fission of $^{236}$U, restricting the model space to 
seniority zero configurations 
of neutrons and protons. 
We particularly focus on 
the distribution of the fission probability. We find that 
it approximately 
follows the chi-squared distribution with the number of degrees of freedom $\nu$ of the order of 1, 
which is consistent 
with the experimental finding. 
We also show 
that $\nu$ corresponds to the number of eigenstates of the many-body Hamiltonian 
whose energy is close to the excitation energy of the system 
and at the same time 
which have 
significant components on both sides of a fission barrier. 
\end{abstract}

\maketitle

\section{introduction}

Heavy compound nuclei decay 
by emitting particles such as neutrons, protons, and alpha particles, 
as well as via fission. 
It has been a custom to describe such 
decays of a compound nucleus 
using a statistical model \cite{frobrich1996,kewpie2}. While a level density is an important microscopic 
input to a statistical model, dynamical calculations based on a many-body Hamiltonian has been rather 
scarce \cite{whitepaper}. 

The purpose of this paper is to develop a microscopic description of decays of a heavy compound 
nucleus, particularly a competition between radiative capture and fission. 
There are many motivations for this. Firstly, in r-process nucleosynthesis, heavy neutron-rich nuclei  
may decay via fission, leading to a fission recycling \cite{cowan2021,eichler2015,goriely2013}. 
Such heavy neutron-rich nuclei are located outside the experimentally known region, and a description 
of fission with a microscopic framework is desirable. Secondly, a neutron separation energy of neutron-rich 
nuclei is so small that a compound nucleus formed in r-process nucleosynthesis will be at relatively low 
excitation energies. One may then question the validity of a statistical model, and thus a microscopic 
approach would be more suitable in that situation. This would be the case also for a barrier-top fission of stable 
nuclei, in which the excitation energy at a saddle of fission barrier will be small due to the presence of a 
barrier. 
An advantage of our model is that a competition between $(n,\gamma)$ and $(n,f)$ processes 
can be described within the same framework. 
Thirdly, because of a rapid increase of computer powers, a large scale calculation can now be 
performed much more easily than before. A microscopic description of fission has been an ultimate goal 
of nuclear physics, and we are now at the stage to tackle it with large scale calculations \cite{whitepaper}.   

In this paper, we propose a novel microscopic approach to low-energy induced fission based on a configuration interaction (CI) 
method. 
This is based on entirely microscopic nucleon interactions except for
input of empirical compound-nucleus properties and the height
of the first fission barrier.
For this purpose, we apply a non-equilibrium Green's function (NEGF) \cite{camsari2023}
to describe decay dynamics. 
This approach 
has been widely utilized to calculate a current and a charge density for 
problems of electron transport in nano-devices \cite{datta_1995,datta_2005}. 
A problem of fission has an analogous feature to this problem, as one has to 
estimate a transmission coefficient for a transition 
from a compound nucleus configuration to a pre-fission 
configuration. This can be viewed as a non-equilibrium current. 

A preliminary calculation with this approach has been published in Ref. \cite{bertsch2023}. 
In that paper, the model space was reduced by considering only neutron seniority-zero configurations 
in $^{236}$U. Moreover, only the dynamics around the first fission barrier was discussed while $^{236}$U is known to have a double humped fission 
barrier. In this paper, we 
shall substantially enlarge the model space, including both neutrons and protons, and also both the first 
and the second fission barriers. 
Such extension of the model space allows a more consistent 
comparison with experimental data. 

With the extended model space, we shall focus particularly on  
the distribution of fission width. 
Decay widths of a compound nucleus are known to follow the chi-squared distribution. 
This distribution is 
characterized by the degrees of freedom $\nu$ \cite{Porter1956,RMP1981}, 
which reflects the number of open exit channels. 
For example, neutron decay widths of very low energy neutrons on
a target with spin zero are well described by 
the chi-squared distribution with $\nu=1$, since there is only a single
($s$-wave) open channel \cite{Liou1972}.   
There are typically many open exit channels for fission decays, reaching the order of 10$^{10}$ for 
low-energy induced fission \cite{Michaudon1972}. 
However, the observed large fluctuations in the fission decay widths require
small values of the fitted $\nu$ parameter.
For example, 
$\nu$ for the $^{235}$U$(n,f)$ reaction was 
found to be 2.3$\pm 1.1$ by fitting the experimental width distribution to the 
chi-squared function \cite{Porter1956}.
In the analysis of more recent and precise data of the same reaction, 
the distributions were well-fitted by the chi-squared distribution with $\nu=2$ \cite{Leal1999}.
For different target nuclei, $^{233}$U and $^{239}$Pu, 
the degrees of freedom have the same order of magnitudes \cite{DERRIEN1993,DERRIEN1994}.

The small values of $\nu$ are explained by assuming that 
fission takes place through a few transition states above a fission barrier. 
Thus the number of degrees of freedom corresponds to the number
of open transition states.
Such transition state hypothesis was  
introduced in the theory of nuclear fission by Bohr and Wheeler \cite{Bohr1939}.
While this has been widely applied to estimate the average fission widths, 
its derivation usually relies on the classical statistical mechanics. 
Even though there have been recent attempts with the 
random matrix approach \cite{Weidenmuller2022,Hagino-TST2023,Weidenmuller2023}, 
its consistency with quantum mechanics has not yet been fully clarified. 
In this paper, we discuss 
the underlying mechanism of the small $\nu$ 
from a miscroscopic point of view. 

The paper is organized as follows. 
In Sec. \ref{model}, we will explain the formulation of our configuration-interaction model. 
In Sec.~\ref{235U}, we will apply the model to the neutron-induced fission of $^{235}$U 
and demonstrate that our model yields a small number of $\nu$. 
We will also discuss its microscopic origin in terms of the behavior of eigenstates of a Hill-Wheeler equation.  
Finally, in Sec.~\ref{summary}, we will summarize the paper and discuss future perspectives.

\section{modeling induced fission reactions \label{model}}

\subsection{Theoretical framework \label{framework}}

We treat a fission process as a transition 
from a compound nucleus state to a pre-fission state 
through 
many-particle many-hole configurations 
along a fission path. 
To this end, we first 
discretize the fission path and 
obtain the local ground state for each point 
based on the constrained density functional theory (DFT) method. 
We then 
construct 
many-particle many-hole configurations 
on top of them. 
Based on the idea of the generator coordinate method (GCM), 
the total wave function is described as 
\begin{equation}
|\Psi\rangle=\int dQ\sum_{\mu}\, f(Q,E_\mu)|Q,E_\mu\rangle, 
\label{GCM}
\end{equation}
where $|i\rangle\equiv|Q,E_\mu \rangle$ represents a Slater determinant characterized by the deformation parameter $Q$ and the excitation 
energy $E_\mu$ from the local ground state. 
Notice that, unlike the usual GCM \cite{ring}, the wave function includes 
not only the local ground states but also many-particle many-hole excited states. 
The GCM kernels are then defined as,
\begin{equation}
H_{i,i'}=\langle i|\hat{H}|i' \rangle=\langle Q,E_\mu|\hat{H}|Q',E_{\mu'} \rangle. 
\label{Hkernel}
\end{equation}
\begin{equation}
N_{i,i'}=\langle i|i'\rangle=\langle Q,E_\mu|Q',E_{\mu'} \rangle,
\label{Nkernel}
\end{equation}

After we construct those kernels based on the constrained DFT  
method, we add imaginary parts $-\frac{i}{2}\Gamma_a$ to the Hamiltonian kernel, Eq. (\ref{Hkernel}), corresponding to the decay width to 
a channel $a$. 
Our model includes a single neutron entrance channel, 
multiple capture channels, and multiple fission channels 
denoted by $\Gamma_{\rm in}$, $\Gamma_{\rm cap}$, and $\Gamma_{\rm fis}$,  respectively. 
Here, $\Gamma_{\rm in}$ and $\Gamma_{\rm cap}$ have components 
in the compound nucleus states, while 
$\Gamma_{\rm fis}$ has components in the pre-fission states. 

The transmission coefficient from a channel $a$ to a 
channel $b$ is computed with 
the Datta formula \cite{datta_1995},
\begin{equation}
T_{a,b}(E)={ \rm Tr}\left[\Gamma_aG(E)\Gamma_bG^\dagger(E)\right], 
\label{Datta}
\end{equation}
where $E$ is the excitation energy of a compound nucleus 
and 
the non-equilibrium Green function $G(E)$ is given by, 
\begin{equation}
G(E)=\left(EN-\left(H-\frac{i}{2}(\Gamma_{\rm in}
+\Gamma_{\rm cap}+\Gamma_{\rm fis})\right)\right)^{-1}.
\end{equation}
Note that we do not need to solve 
the Hill-Wheeler equation if the Green function is 
constructed with a matrix inversion 
technique \cite{Bertsch2022}. 
In our model, the input channel $a$ corresponds to a neutron channel, while the output channel $b$ is either 
a fission channel or a capture channel.

\subsection{Chi-squared distribution and its degrees of freedom}

In this paper, we will discuss a fluctuation of 
the transmission coefficients for the fission channel, 
$T_{{\rm in, fis}}$, and its relation to the chi-squared 
distribution. 
Here, 
the chi-squared distribution $P_{\nu}(x)$ is defined 
as, 
\begin{equation}
P_\nu(x)=\frac{\nu}{2\Gamma(\nu/2)}\left(\frac{\nu x}{2}\right)^{\nu/2-1}e^{-\nu x/2}.
\label{P_nu}
\end{equation}
The parameter $\nu$ is referred to as degrees of freedom and $\Gamma$ is the Gamma function. 
Empirically, the decay width of compound nucleus states 
is known to closely follow 
the chi-squared distribution in many cases \cite{Porter1956}.

Note that the transmission coefficient obtained with Eq.(\ref{Datta}) includes 
the fluctuations of both the input channel $a$ and the output channel $b$. 
Therefore, we use the fission probability \cite{Hagino2021}, 
\begin{equation}
\label{pfis}
P_{\rm fis}\equiv T_{{\rm in,fis}}/T_{\rm in}\sim T_{{\rm in,fis}}/(T_{{\rm in,fis}}+T_{{\rm in,cap}}),    
\end{equation}
rather than $T_{{\rm in,fis}}$ itself. Here, the relation $T_n\simeq T_{{\rm in,fis}}+T_{{\rm in,cap}}$ 
is derived from the unitarity of the $S$-matrix and its validity has been confirmed in Appendix in Ref. \cite{bertsch2023}.
An advantage to use $P_{\rm fis}$ is that the fluctuation of the input channel is cancelled out in it between the denominator and the 
numerator. 

\section{Application \lowercase{to} $^{235}{\rm U}(n,f)$ \label{235U}}
\subsection{A setup of the model}

Let us now apply the theoretical framework 
to a neutron-induced fission reaction, $^{235}$U$(n,f)$. 
To this end, 
we construct 
the GCM basis functions, $|Q,E_\mu\rangle$, 
with the density-constranied DFT calculation, 
assuming that the fission path is given by the mass 
quadrupole moment, $Q_{20}\equiv Q_2$, with axial symmetry.  
As a DFT solver, we employ {\tt Skyax} \cite{Reinhard2021},  in which the Kohn-Sham equation is solved in the cylindrical coordinate space.
As an energy functional, we use 
a Skyrme functional with the 
UNEDF1 parameter set \cite{UNEDF1}, 
which has an effective mass close to one and thus 
is suitable to reproduce a reasonable level density of 
excited nuclei. 
Note that the reference states are Slater determinants:
a pairing interaction is included later as a residual 
interaction between the states. 

\begin{figure}
\includegraphics[width=8.6cm]{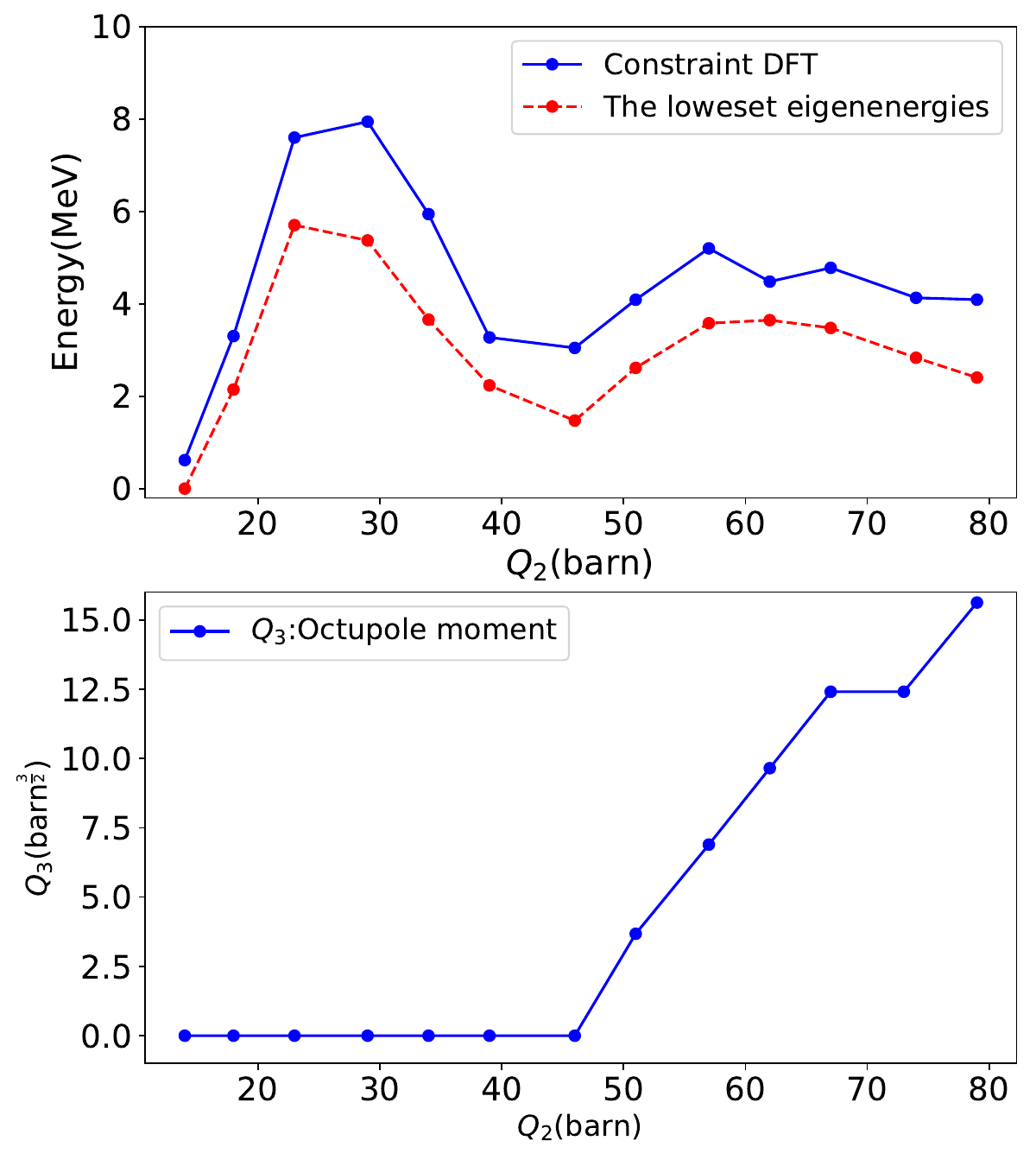}
\caption{
(The upper panel) 
The fission barrier of $^{236}$U along the 
fission path defined by the mass quadrupole moment, $Q_2$.  
The blue solid line shows the energies of the 
local ground states obtained with the constrained DFT calculation. It is scaled by a factor of $0.71$.
The red dashed line shows the lowest eigenvalues obtained by diagonalizing each $Q$-block after scaling the blue solid 
line.
The origin of the energy is set at the lowest eigenvalue at $Q_2=14$ b. 
(The lower panel) The octupole moment $Q_3$ in $^{236}$U along the fission path. 
}
\label{barrier}
\end{figure}

The fission path is discretized 
with a criterion that 
the overlap of the local ground states between the nearest neighbors is ${\cal N}\sim e^{-1}$ \cite{bertsch2023, Bertsch2022}. 
We extend the maximum value of $Q$ up to around 80 b so that both the first 
and the second fission barriers are covered. 
The criterion for the discretization leads to 13 blocks from $Q=14$ b to $Q=79$ b. 
The potential energy curve for 
fission of $^{236}{\rm U}$ 
is shown 
in 
the upper panel of 
Fig. \ref{barrier} by the blue solid line 
as a function of the quadrupole moment $Q_2$, together with the octupole moment $Q_3$ 
shown in the lower panel. 
In this calculation, 
the ground state is located at $Q_2=14$ b. There are two fission barriers, 
the first fission barrier around $Q_2=30$ b 
and the second barrier around $Q_2=60$ b. 
The fission path is along the mass symmetric path 
up to the first barrier, and it extends to the 
mass asymmetric path going through the second barrier, 
as is indicated in the lower panel of Fig. \ref{barrier}. 

In the previous work \cite{bertsch2023}, 
the many-body configurations were constructed solely with neutron excitations  
up to 4 MeV. In contrast,  
in this paper we extend the model space 
and take both neutron and 
proton excitations up to 5 MeV. 
Following  
Ref. \cite{bertsch2023}, we shall   
take into account only seniority-zero configurations, that is, those without 
broken pairs. As a result, the dimension of the Hamiltonian kernel becomes the order of $6\times10^4$.
We call a sub-block in the Hamiltonian kernel for 
each $Q$ a $Q$-block, and 
the dimension of each $Q$-block is summarized in Table \ref{dim1}.

\begin{table*}[htb] 
\begin{tabular}{c|cccccccccccccc}
\hline 
\hline 
$Q_2$(barn)  & 14& 18&  23 & 29 & 34 & 39 & 46 & 51 & 57 &62 & 67 & 74 & 79 &83 \\
\hline 
 dim.& $N_{
 \rm GOE}$&2520&  9794 & 15088 & 11577 & 2774 & 2940 & 3021 & 3150 & 2196 & 3752 & 2871 & 4420 &$N_{\rm GOE}$\\
\hline 
\hline 
\end{tabular}
\caption{The dimension of each $Q$-block 
for fission of $^{236}$U. 
\label{dim1}}
\end{table*} 

In the calculation of the Hamiltonian kernel, 
the residual interactions between configurations includes a monopole pairing
component
\begin{equation}
  H_{\rm pair}=-G\sum_{i\neq j}a^{\dagger}_{i}a^{\dagger}_{\bar{i}}a_{\bar{j}}a_{j},  
  \label{eq:H_pair}
\end{equation}
and a diabatic component \cite{Hagino2022}, 
\begin{eqnarray}
\frac{\langle Q,E_\mu|v_{db}|Q',E_{\mu'}\rangle}{\langle Q,E_\mu|Q',E_{\mu'}\rangle}
&=&\frac{E(Q,E_\mu)+E(Q',E_{\mu'})}{2} \nonumber \\
&&+h_2{\rm ln}(\langle Q,E_\mu|Q',E_{\mu'}\rangle).  
\end{eqnarray}
The $\bar{i}$ in Eq. (\ref{eq:H_pair}) denotes the time-reversal state of $i$. 
The diabatic interaction acts only between 
the diabatically connected configurations, 
$|Q,E_{\mu}\rangle$ and $|Q',E_{\mu'}\rangle$
\cite{Hagino2022}.
We take $G=0.16$ MeV and $h_2=1.5$ MeV. 
The value of $G$ is determined to reproduce the excitation energy of the first excited $0^+$ state of $^{236}$U
within the model space so specified \cite{bertsch2023} 
and the value of $h_2$ is the same as the one used in Ref. \cite{bertsch2023}.

The red dashed line in the upper panel of Fig. \ref{barrier} shows the potential energy curve connecting the lowest eigenvalue for each $Q$-block.
To reproduce 
the experimentally determined barrier height of $5.7$ MeV \cite{Lindgren1978}, 
we have introduced a multiplicative factor of 0.71 
to the solid line and then diagonalized the Hamiltonian for each 
$Q$-block. 
At least for the first barrier, the overestimation of the barrier height 
may be partly 
attributed to the absence of the triaxial deformation \cite{Staszczak2009}.
We have confirmed that the results shown below remained qualitatively the same even if the 
rescaling was applied only to the first barrier. 

As the dimension is still large for the $Q$-block at $Q_2=14$ b as well as 
the $Q$-block right after $Q_2=79$ b, 
we follow the previous calculation \cite{bertsch2023}  
and replace those 
with a random matrices sampled 
from the Gaussian Orthogonal Ensemble (GOE).
We set the central energy of the matrices to be the same as the excitation energy, $E$.
In addition to the central energy, 
the GOE is characterized 
by the root-mean-square of the matrix elements 
$\langle v^2 \rangle^{1/2}$ and the matrix dimension 
$N_{\rm GOE}$. 
These parameters are related 
with the level density at the center of the distribution,  
$\rho_0=N_{\rm GOE}^{1/2}/\pi\langle v^2\rangle^{1/2}$ \cite{WeidenmullerReview}.  
In our calculations, we set $\rho_0=31.8$ MeV$^{-1}$ 
\cite{bertsch2023}
and $N_{\rm GOE}=1000$. 
Notice that the configurations are strongly mixed 
after the diagonalization the GOE matrix. Therefore, 
a neutron can be emitted from any configuration  
within the GOE space, even if
a sigle neutron channel is considered in $\Gamma_{\rm in}$.
See Appendix \ref{GOE} for general properties of a GOE Hamiltonian including decay widths. 

The left-most GOE-matrix represents the compound nucleus states having decay probabilities corresponding to neutron emission and $\gamma$ decay.
Therefore we add imaginary matrices $-\frac{i}{2}\Gamma_{\rm in}$ and $-\frac{i}{2}\Gamma_{\rm cap}$ to the Hamiltonian kernel.
The matrix $\Gamma_{\rm in}$ has the value $\gamma_{\rm in}$ in the first diagonal component 
and all the other elements are zero, while 
$\Gamma_{\rm cap}$ has the following structure,
\begin{equation}
\Gamma_{\rm cap}  = \left(\begin{matrix}
     \tilde{\Gamma}_{\rm cap} & 0 &  & \cr
     0& 0 &  &  \cr
      &  & \ddots &  \cr
      &  &  &0 \cr
      \end{matrix}\right), 
\end{equation}
where $\tilde{\Gamma}_{\rm cap}=\gamma_{\rm cap}I$, with $I$ being the unit matrix with the dimension 
$N_{\rm GOE}$. 

Following 
the Appendix in Ref. \cite{bertsch2023}, we set $\gamma_{\rm in}=0.01$ MeV and $\gamma_{\rm cap}=0.00125$ MeV, respectively.
Those width parameters are chosen with the help of compound-nucleus
phenomenology through the formula relating the transmission coeffient
into the compound nucleus and the average widths of compound-nucleus
states, 
\begin{equation}
T_k = 2 \pi \langle\gamma\rangle \rho.
\end{equation}

The right-most GOE-matrix represents pre-fission configurations. 
We therefore add to it an imaginary decay  matrix $-\frac{i}{2}\Gamma_{{\rm fis}}$ for a fission decay.
The structure of $\Gamma_{\rm fis}$ is given by 
\begin{equation}
\Gamma_{\rm fis}  = \left(\begin{matrix}
     0 &  &  & \cr
     & \ddots &  &  \cr
      &  & 0 & 0 \cr
      &  & 0 &\tilde{\Gamma}_{\rm fis} \cr
      \end{matrix}\right), 
\end{equation}
where $\tilde{\Gamma}_{\rm fis}=\gamma_{\rm fis}I$.
It has been found that transmission coefficients 
are insensitive to the value of $\gamma_{{\rm fis}}$ 
\cite{bertsch2023,Uzawa2023}, 
and we set $\gamma_{{\rm fis}}$ arbitrarily to be 
$0.015$ MeV.

Neglecting the couplings 
between the next-to-the nearest neighboring $Q$-blocks \footnote{This
approximation has been analyzed in 
Ref. \cite{BH2024}.}, 
the resultant Hamiltonian matrix has the following structure
\begin{equation}
\label{3D-H0}
H  = \left(\begin{matrix}
     \tilde{H}^{(L)}_{\rm GOE} & (V^{(L)})^T &  & & & \cr
     V^{(L)} & H_1 & V_{1,2} &  & \text{\large{\textit{O}}}& \cr
      & V_{2,1} & H_2 & V_{2,3} &  & \cr
      &  &  &\ddots &  & \cr
      \text{\large{\textit{O}}}&  & & V_{11,12}& H_{12} &(V^{(R)})^T \cr
      &  & & & V^{(R)} &\tilde{H}^{(R)}_{\rm GOE} \cr
           \end{matrix}\right), 
\end{equation}
where $O$ is the zero matrix, and 
$\tilde{H}^{(L)}_{\rm GOE}$ and $\tilde{H}^{(R)}_{\rm GOE}$ denote the GOE random matrices
including decay widths.
$H_k$ represents the matrix elements for the configurations at specific $Q_k$. 
$V_{k,k'}$ denotes off-diagonal block components between neighboring configurations.
We assume that 
the matrix elements of $V^{(L)}$ and $V^{(R)}$ also follow a Gaussian distribution, with r.m.s strengths set to be  $\sqrt{\langle v_a^2 \rangle}=0.02$ MeV and $\sqrt{\langle v_b^2 \rangle}=0.03$ MeV, respectively. Those order of magnitude 
may be justified as follows. 
The present calculation with the UNEDF1 parameter set yields the  
level density of $\rho_{\rm tot}=3.87\times 10^5$ MeV$^{-1}$ for $K^\pi=0^+$ configurations, where 
$K$ is the spin projection onto the symmetry axis, at the excitation energy $E=6.5$ MeV. 
On the other hand, if the configurations are restricted only to the seniority zero, the 
level density is $\rho_{\nu=0}=220$ MeV$^{-1}$ at the same excitation energy. 
If one scales the strength of the diabatic interaction according to the level 
densities, the strength of a residual interaction is estimated to be 
$v=h_2\sqrt{\rho_{\nu=0}/\rho_{\rm tot}}\sim 0.036$ MeV for $h_2=1.5$ MeV. 
This is close to the values of $v_a$ and $v_b$ which we employ.

\begin{figure}
\includegraphics[width=4cm]{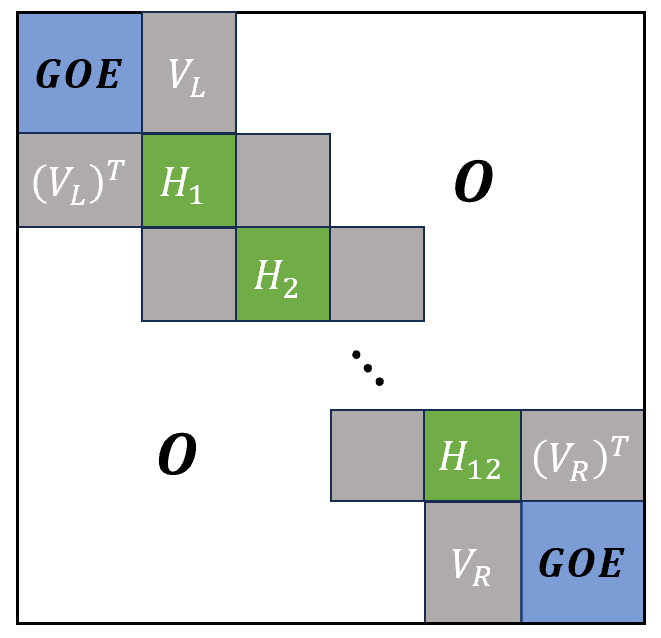}
\caption{
A schematic illustration of the Hamiltonian 
matrix. 
}
\label{H}
\end{figure}

The overlap kernel has a similar structure,
\begin{equation}
N  = \left(\begin{matrix}
     I^{(L)} &O  & & &\text{\large{\textit{O}}} \cr
      O& I_1 & S_{1,2} &  & & \cr
      & S_{2,1} & I_2 & S_{2,3} &  & \cr
      &  &  &\ddots &  & \cr
      \text{\large{\textit{O}}}&  & & S_{11,12}& I_{12} &O \cr
      &  & & &  O&I^{(R)} \cr
           \end{matrix}\right), 
\end{equation}
where $I$ represents the identity matrix. 
$S_{k,k'}$ represents the overlap between neighboring $Q$-block configurations. As in the Hamiltonian 
matrix, we ignore the overlap between the next-to-the nearest 
neighboring configurations. 
With this simplification, the matrix $(EN-H)$ becomes block tri-diagonal and the inversion matrix $G(E)$ can be efficiently calculated with the method presented in 
Ref. \cite{Petersen2008}. See Fig. \ref{H} for a schematic illustration of the Hamiltonian matrix.

\subsection{The transmission coefficients}

\begin{figure}
\includegraphics[width=7cm]{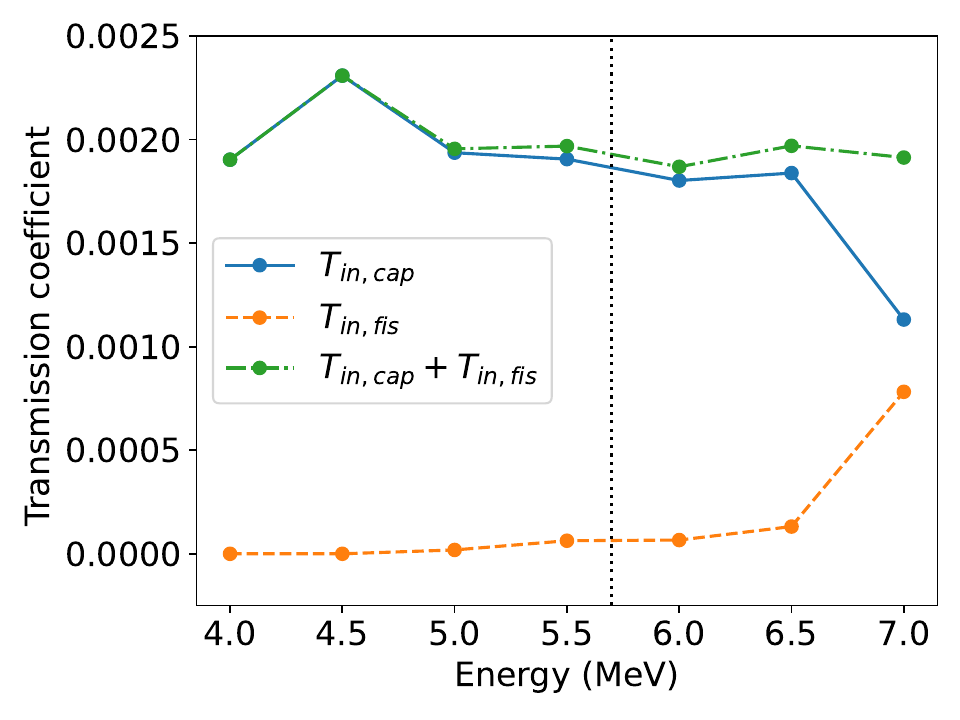}
\caption{The averaged transmission coefficients for capture $\langle T_{{\rm in},{\rm cap}}(E)\rangle$ 
(the blue solid line) 
and for fission $\langle T_{{\rm in},{\rm fis}}(E)\rangle$ 
(the orange dashed line) as a function of the excitation energy $E$. 
The sum of these transmission coefficients is also plotted with the green dot-dashed line.
The vertical dotted line shows the height of the fission barrier located at $5.7$ MeV.
}
\label{ratio}
\end{figure}

Let us now numerically evaluate the transmission coefficients, $T_{{\rm in},{\rm cap}}$ and $T_{{\rm in},{\rm fis}}$.
Experimentally, decay widths are measured within an energy resolution.   
We thus introduce an energy average, 
\begin{equation}
\langle T_{{\rm in},a}(E)\rangle=\frac{1}{\Delta E}\int^{E+\Delta E/2}_{E-\Delta E/2} dE' T_{{\rm in},a}(E'), 
\label{average}
\end{equation}
where $\Delta E$ is an energy interval. We take $\Delta E=0.25$ MeV, which satisfies the condition $\Delta E \gg 1/\rho_0$.
Furthermore,  
we take an ensemble average with 100 samples 
of the transmission coefficients. 
Fig. \ref{ratio} shows 
the energy dependence of the transmission coefficients so obtained 
for the capture (the solid line) and the fission (the dashed line). 
$\langle T_{{\rm in},{\rm fis}}(E)\rangle$ increases as the excitation energy increases, 
while $\langle T_{{\rm in},{\rm cap}}(E)\rangle$ decreases 
because the total reaction probability is approximately conserved (see the dot-dashed line). 
At $E=6.5$ MeV, which is close to the neutron separation energy of $^{236}$U ($S_n=6.536$ MeV) \cite{Leal1999}, 
the fission-to-capture branching ratio, $\alpha^{-1}\equiv \langle T_{{\rm in},{\rm fis}}\rangle/\langle T_{{\rm in},{\rm cap}}\rangle$, is 
$0.071$ in this calculation. 
Even though this value is still reasonable, it underestimates the empirical value, $\alpha^{-1}\simeq3$ \cite{Moore1978}, by a 
factor of about 40. One could increase the values of $v_a$ and $v_b$ to obtain a more reasonable 
branching ratio. However, we have found that the fluctuation of $T_{{\rm in},{\rm fis}}(E)$ then largely deviates from the chi-squared 
distribution, which is inconsistent with experimental findings. Since we employ the 
justifiable values of $v_a$ and $v_b$, this clearly indicates that one needs to further increase the model space to reproduce the empirical branching ratio.  
In fact, 
it would be expected that the agreement with the experimental branching ratio is 
improved by including seniority non-zero configurations and a proton-neutron random interaction which acts on that space 
\cite{Uzawa2023,Bush1992}.

\subsection{Distribution of $P_{{\rm fis}}$}

\begin{figure}
\includegraphics[width=8.6cm]{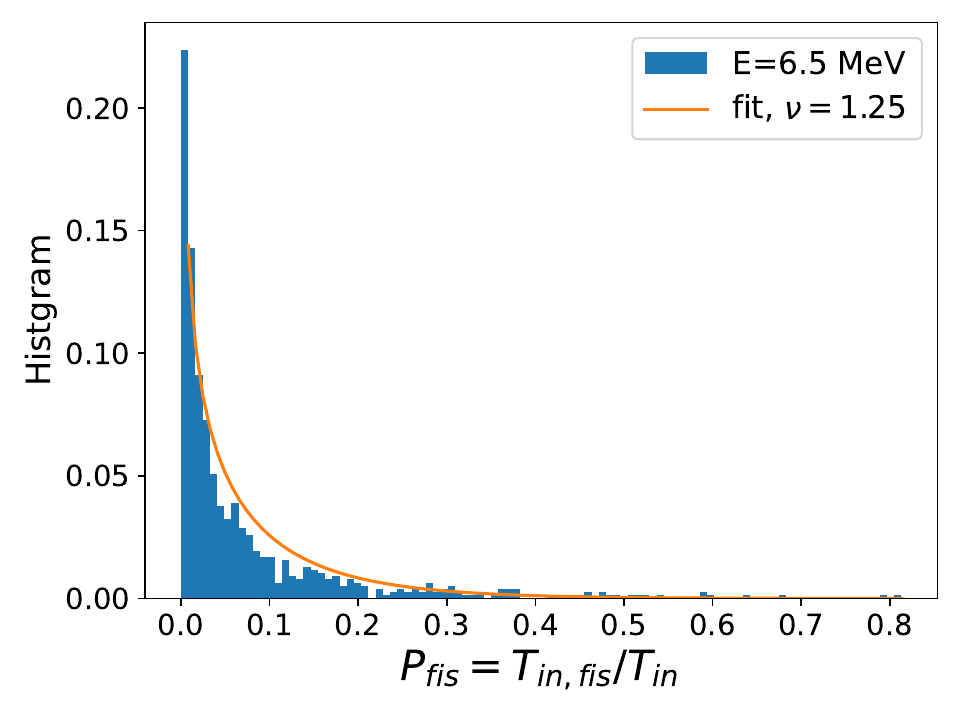}
\caption{Distributions of $P_{{\rm fis}}(E)$ for 1000 samples at $E=6.5$ MeV. 
The orange solid line shows a chi-squared distribution with $\nu$ determined by the maximum likelihood fit. 
The value of $\nu$ is shown in the inset. 
}
\label{hist1}
\end{figure}

\begin{figure}
\includegraphics[width=8.6cm]{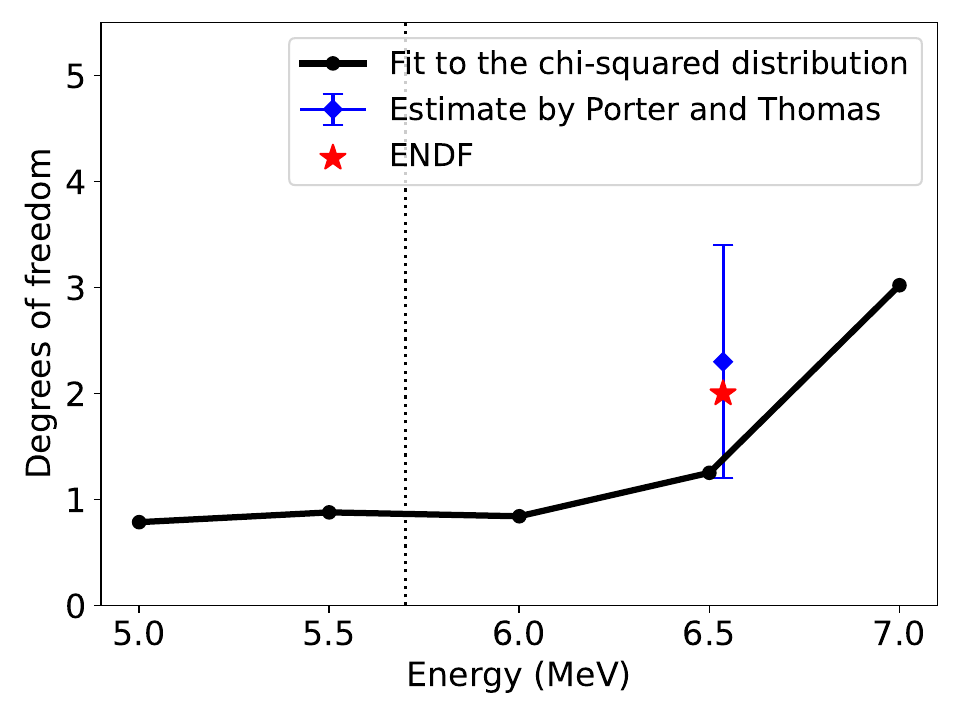}
\caption{
The number of degrees of the freedom $\nu$  
obtained by fitting the distribution of transmission coefficients for fission 
to the chi-squared distribution. 
The blue diamond is the empirical estimate of $\nu$ in Ref. \cite{Porter1956}, while 
the star represents the data from ENDF/B-VIII.0 \cite{Brown20181short,osti_631239}. 
The vertical dotted line denotes 
the height of the fission barrier.}
\label{nu1}
\end{figure}

An important quantity for induced fission is the number of degree of freedom, 
which is related to the effective number of decay channels.
In order to study this,
we examine the fluctuation of the fission channel $P_{\rm fis}$ in Eq.(\ref{pfis}).  
To this end, we fit the distribution of $P_{\rm fis}$ generated with 1000 samples for a specific excitation energy $E$ with 
the chi-squared function defined by Eq.(\ref{P_nu}).
The distribution of $ P_{\rm fis}$ at $E=6.5$ MeV is shown in Fig. \ref{hist1}, while 
extracted values of $\nu$ are shown in Fig. \ref{nu1}.
It is remarkable that the extracted $\nu$ is much smaller than the number of fission channels, that is, 
$N_{\rm GOE}=1000$ in this calculation. 
This is consistent with the picture of transition state theory \cite{Bohr1939, truhlar1983, truhlar1996, Mills1994, miller1974, Laidler1983, marcus1951, marcus1952} 
and our model yields it naturally even though 
we do not introduce apriori 
any assumption used in it \cite{bertsch2021}. 
The value near $E=6.5$ MeV, $\nu=1.25$, is close to the original estimate 
of Porter and Thomas \cite{Porter1956}, ($\nu=2.3\pm1.1$ at $E$=6.536 MeV),
as well as a recent estimate based on evaulated cross-section data
\cite{Brown20181short}, even though the calculation somewhat underestimates the empirical values
\footnote{We expect that the agreement is improved if seniority non-zero configurations are 
taken into account in the model space.}.

Incidentally, the result shown in Fig. \ref{nu1} is consistent with Fig. 3 in Ref.\cite{FUSION}, in which the degrees of freedom $\nu$ was extracted based on the rank of $\Gamma_{\rm eff}$ defined by Eq. (\ref{eHamiltonian}) below.
The consistency of the results obtained with the different approaches strongly supports the validity of our finding of small $\nu$.

One can see in Fig. \ref{hist1} that the distribution of $P_{\rm fis}$ approximately follows the chi-squared distribution; however, the agreement is not perfect. We will discuss a possible origin for the deviation in the next subsection based on the effective Hamiltonian approach.

\subsection{Effective Hamiltonian %$H_{\rm eff}$
for a compound nucleus configuration
\label{eff}}

\begin{figure}
\includegraphics[width=8.6cm]{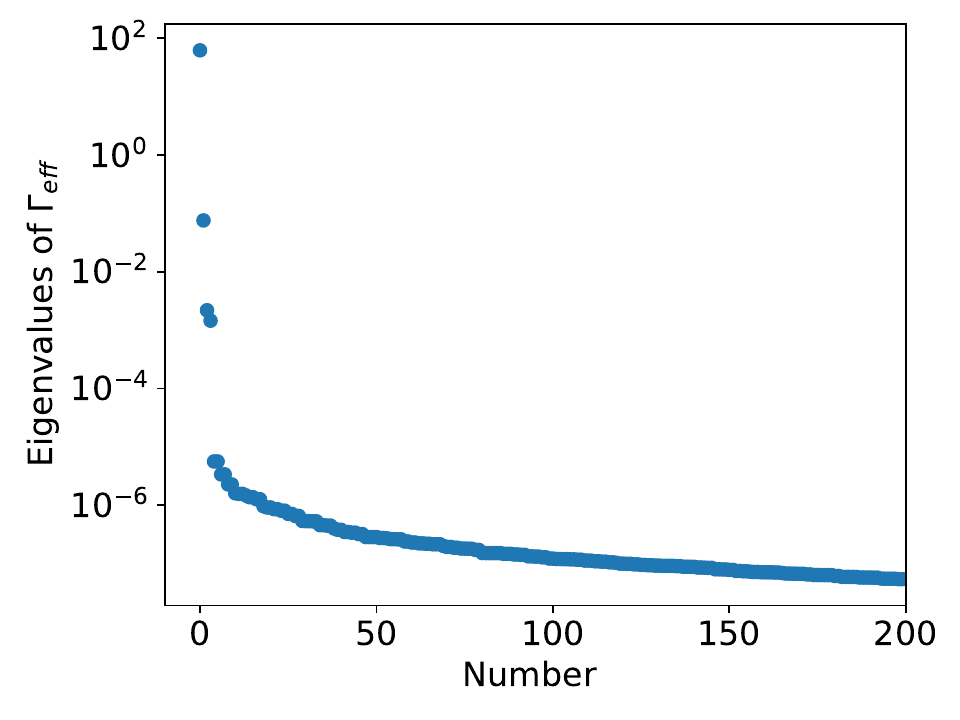}
\caption{Eigenvalues of the matrix $\Gamma_{\rm eff}$ at $E=6.5$ MeV defined by Eq. (\ref{eHamiltonian}) 
for a typical sample. 
The dimension of $\Gamma_{\rm eff}$ is $1000$ and the first 200 eigenvalues are plotted in descending order.
}
\label{distofEV2}
\end{figure}

The fact that the fission probability behaves closely to the chi-squared distribution   
originates 
from the properties of the GOE matrix (see Appendix \ref{GOE}). 
To discuss how it arises, we here construct 
an effective Hamiltonian for the compound nucleus configurations by 
eliminating the other space. 
As we would like to discuss the fluctuation of the fission width, in this subsection, we set $\Gamma_{\rm in}=\Gamma_{\rm cap}=0$ 
and consider the width matrix only for the fission channel. 
Let us write the Hamiltonian, Eq. (\ref{3D-H0}), as 
\begin{equation}
H  = \left(\begin{matrix}
     H^{(L)}_{\rm GOE} & (\vec{V}^{(L)})^T \cr
     \vec{V}^{(L)} & H_Q 
           \end{matrix}\right), 
\end{equation}
where $\vec{V}^{(L)}$ is defined as  $\vec{V}^{(L)}=(V^{(L)},O,O,\cdots,O)^T$.
%$H_Q$ represents the Hamiltonian matrix 
%excluding the first $N_{\rm GOE}$ rows and colums 
%from the original Hamiltonian matrix, $H$. 
We define $N_Q$ in a similar way for the overlap kernel, $N$. 
The effective Hamiltonian for the space of $H^{(L)}_{\rm GOE}$ can then be 
constructed as  
\begin{align}
H_{\rm eff}(E)&=
H^{(L)}_{\rm GOE}-\vec{V}^{(L)}(H_Q-EN_Q)^{-1}(\vec{V}^{(L)})^T \notag\\
&\equiv H^{(L)}_{\rm GOE}+\Delta(E)-i\Gamma_{\rm eff}(E)/2, 
\label{eHamiltonian}
\end{align}
where $H^{(L)}_{\rm GOE}+\Delta(E)$ and $-\Gamma_{\rm eff}(E)/2$ are the real and the imaginary 
parts of the effective Hamiltonian, respectively. 
$\Delta(E)$ serves as an energy shift and $\Gamma_{\rm eff}(E)$ corresponds to the fission width for the compound states. 
Notice that the width matrices, $\Gamma_{\rm cap}$, $\Gamma_{\rm fis}$, 
and $\Gamma_{\rm in}$, have the same diagonal structure to Eq. (\ref{Gamma}), but this 
may not be the case in $\Gamma_{\rm eff}(E)$.

If $\Delta(E)$ was zero, the real part of $H_{\rm eff}$ became a GOE matrix itself, 
and the degrees of freedom of the exit channel was estimated by \cite{Miller1990}
\begin{equation}
    \nu=\frac{{\rm Tr}[\Gamma_{\rm eff}]^2}{{\rm Tr}[(\Gamma_{\rm eff})^2]}.
    \label{Tr}
\end{equation}
If we apply this formula to our calculation, we obtain $\nu=1.00$ at $E=6.5$ MeV, which is consistent with the result shown in Fig. \ref{nu1}.
The eigenvalues of $\Gamma_{\rm eff}$ at $E=6.5$ MeV are plotted in Fig. \ref{distofEV2} for a specific random seed.
In our model, 
the dimension of $\Gamma_{\rm eff}$ is $N_{\rm GOE}=1000$ and there are 1000 eigenvalues for each ensemble. 
One can notice that 
there exists only one large eigenvalue and the remainders are negligibly small as compared to it. 
Naturally the value of $\nu$ becomes close to 1 if Eq. (\ref{Tr}) is applied. We have confirmed that  
this is the case for all the samples which we study in this paper (see Appendix \ref{App:eigenvalues}). 

In reality, a finite $\Delta(E)$ makes the real part of $H_{\rm eff}$ deviate from a pure GOE matrix, and the distribution is 
also perturbed from a pure chi-squared distribution.
In our setup, the effect of $\Delta(E)$ is small and the distribution still follows 
approximately a chi-squared distribution (see Fig. \ref{hist1}).

\subsection{Discussion\label{discussion}}

In the previous subsection, 
we have 
investigated the eigenvalues of the decay matrix, $\Gamma_{\rm eff}$, and 
demonstrated that a fission width has small degrees of freedom. 
In order to understand it microscopically, 
let us go back to the Datta formula, Eq. (\ref{Datta}). 
With the setup of our model for $\Gamma_{\rm in}$ and $\Gamma_{\rm fis}$, this formula reads, 
\begin{equation}
T_{{\rm in,fis}}=\gamma_{\rm in}\gamma_{{\rm fis}}\sum_{j\in {\rm fis}}|G_{1,j}|^2.
\label{Tnf}
\end{equation}
Here the neutron channel $n=1$ represents a specific configuration in the left-end GOE 
and the fission channel includes all configurations in the right-end GOE.
We then perform a spectrum decomposition of $G(E)$ as
\footnote{
This is in contrast to the Appendix of Ref. \cite{Porter1956}, 
in which a decaying wave function was decomposed into transition states. 
},
\begin{equation}
G_{ij}(E)=\sum_\lambda f_i^{(\lambda)}\,\frac{1}{E-\tilde{E}_\lambda}\,
(f_{j}^{(\lambda)})^*,
\label{SD}
\end{equation}
where $f_\mu^{(\lambda)}$ is a solution of the generalized eigenvalue problem with the GCM kernels in Eqs. (\ref{Nkernel}) 
and (\ref{Hkernel}), satisfying 
\begin{equation}
    \sum_{j}(H-E_\lambda N)_{ij}f_{j}^{(\lambda)}=0. 
\end{equation}
Notice that 
$f_j^{(\lambda)}$ with $j=(Q,E_\mu)$ is equivalent to 
the GCM weight function $f_\lambda(Q,E_\mu)$ defined by Eq. (\ref{GCM}). 
$\tilde{E}_\lambda$ in Eq. (\ref{SD}) is defined as 
$\tilde{E}_\lambda=E_\lambda-\frac{i}{2}\Gamma_\lambda$, where 
 $\Gamma_\lambda$ is given by 
\begin{equation}
\Gamma_{\lambda}=\sum_{i,j}
(f_i^{(\lambda)})^*(\Gamma_{\rm in}+\Gamma_{{\rm cap}}+\Gamma_{{\rm fis}})_{ij}
f_{j}^{(\lambda)}.
\label{gamma_perturbation}
\end{equation}
Notice that for simplicity the decay matrices in the Green function are treated perturbatively. 
Substituting Eq.(\ref{SD}) into Eq.(\ref{Tnf}), we obtain 
\begin{align}
T_{{\rm in,fis}}
&=\gamma_{\rm in}\gamma_{{\rm fis}}\sum_{\lambda} 
\frac{|f_1^{(\lambda)}|^2}{(E-E_\lambda)^2+(\Gamma_\lambda/2)^2} 
\sum_{j\in {\rm fis}} |f_j^{(\lambda)}|^2 \notag\\
&+\gamma_{\rm in}\gamma_{{\rm fis}}\sum_{\lambda\neq\lambda'}
\sum_{j\in {\rm fis}}
\frac{f_1^{(\lambda)}f_1^{(\lambda')*} 
f_j^{(\lambda)}f_j^{(\lambda')*} }{(E-\tilde{E}_\lambda)(E-\tilde{E}_\lambda')^*}.
\label{T_decomposition}
\end{align}

We then take an ensemble average of $T_{{\rm in, fis}}$. 
To this end, we notice that 
$f_k^{(\lambda)}$ approximately follows a Gaussian distribution, 
and they are uncorrelated with the eigenvalues $\tilde{E}_\lambda$ \cite{bertsch2021} 
when $k$ is for the neutron and the fission channels. 
As explained in Appendix \ref{GOE}, amplitudes of GOE eigenstates follow in general a Gaussian distribution, 
and this property is expected to be conserved in our model as long as the couplings 
between the GOE matrices and the bridge Hamiltonian are not too strong. 
As we have discussed in Sec. \ref{eff}, we have confirmed that this is the case 
for the coupling strengths which we employ, that is, $\sqrt{\langle v_a^2 \rangle}=0.02$ MeV 
and $\sqrt{\langle v_b^2 \rangle}=0.03$ MeV. 

As a consequence, the second term in Eq. (\ref{T_decomposition}) vanishes and 
one can take an ensemble average separately for the three factors in the first term in Eq. (\ref{T_decomposition}).  
The ensemble averaged transmission coefficients for fission then reads
\begin{align}
\langle T_{{\rm in, fis}}(E)\rangle&= \gamma_{\rm in}\gamma_{{\rm fis}}\sum_{\lambda}\langle |f_\lambda(Q_L,E_n)|^2\rangle\notag \\
&\times\left\langle\frac{1}{(E-E_\lambda)^2+(\Gamma_\lambda/2)^2}\right\rangle 
\left\langle \sum_{\mu} |f_\lambda(Q_R,E_\mu)|^2\right\rangle,  
\label{T_decomposition2}
\end{align}
where $Q_L$ and $Q_R$ denote the left-most and the right-most configurations, respectively. 
In this way, $T_{in,{\rm fis}}$ is decomposed into a contribution of each GCM eigenmode, $\lambda$.

\begin{figure}[tb]
\includegraphics[width=8.6cm]{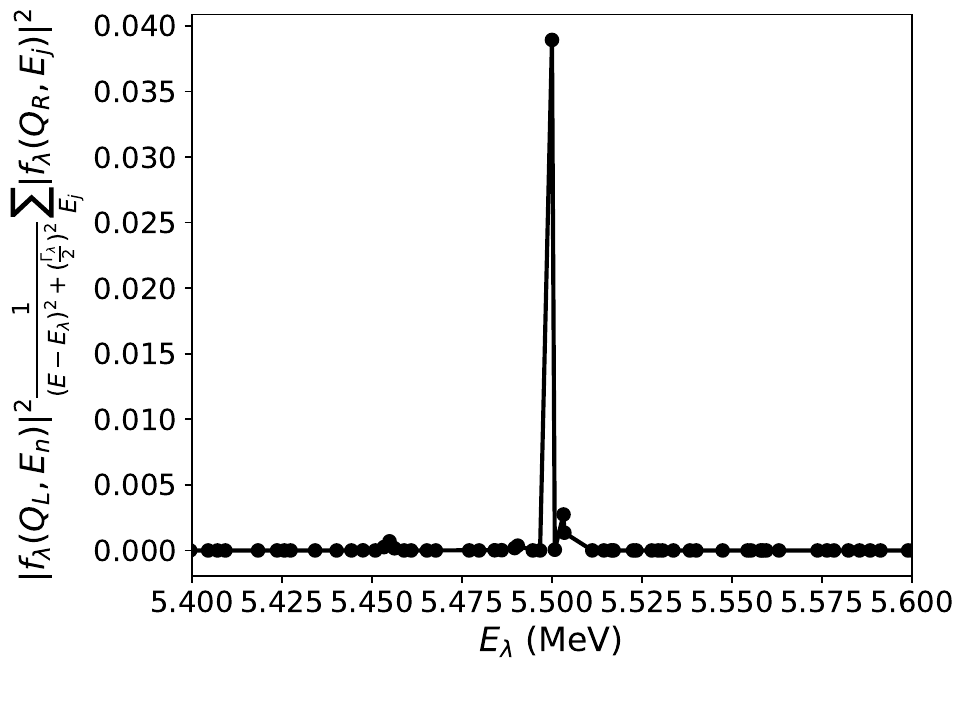}
\caption{The spectrum decomposition of the transmission coefficient for fission at $E=5.5$ MeV 
defined by Eq. (\ref{T_decomposition2}). 
}
\label{Dattafull}
\end{figure}

In order to investigate how many
eigenmodes contribute to the transmission coefficient, 
Fig. \ref{Dattafull} plots the contribution of each eigenmodes 
for $E=5.5$ MeV 
as a function of $E_\lambda$. 
The Breit-Wigner term acts as an energy-window, 
and the eigenmodes $\lambda$ 
contribute significantly to $T_{in,{\rm fis}}$ only when  
the eigenenergy $E_\lambda$ is within 
the range $(E-\Gamma_\lambda/2,E+\Gamma_\lambda/2)$. 
Table \ref{table:components} shows the breakdown of each term 
in Eq. (\ref{T_decomposition2}) 
for five eigenstates around the dominant eigenmode. 
One can see that the components both at the left-most and the right-most $Q$ are relatively 
large for the dominant eigenmode as compared to those for the other eigenmodes.  
This is a necessary condition to have a large transmission coefficient, as is evident from 
Eq. (\ref{T_decomposition2}). 

\begin{table*}[htb] 
\begin{tabular}{c|c|c|c|c|c|c}
\hline 
\hline 
$E_\lambda$ (MeV)  & $|f_\lambda(Q_L,E_n)|^2$& $\frac{1}{(E-E_\lambda)^2+(\Gamma_\lambda/2)^2}$ & $\sum_{\mu}|f_\lambda(Q_R,E_\mu)|^2$ & %$\frac{|f_\lambda(Q_L,E_n)|^2\sum_{\mu}|f_\lambda(Q_R,E_\mu)|^2}{(E-E_\lambda)^2+(\Gamma_\lambda/2)^2}$
the product 
& $\Gamma_\lambda$ (MeV)\\
\hline 
 $5.4946$&$2.56\times10^{-9}$ &$3.51\times10^{4}$   &  $2.82\times10^{-2}$& 2.58$\times 10^{-6}$ & 4.23$\times 10^{-4}$\\
\hline
 $5.4969$&$1.30\times10^{-5}$ & $9.04\times10^{4}$  &  $2.26\times10^{-6}$ & 2.68$\times 10^{-6}$ & 5.68$\times 10^{-4}$\\
\hline 
 $5.4999$&$1.17\times10^{-7}$ & $6.22\times10^{6}$  &  $5.34\times10^{-2}$ & 3.89$\times 10^{-2}$& 8.02$\times 10^{-4}$ \\

\hline 
 $5.5008$&$9.54\times10^{-9}$ & $1.66\times10^{6}$  &  $3.59\times10^{-3}$& 5.68$\times 10^{-5}$& 5.39$\times 10^{-5}$ \\

\hline 
 $5.5032$&$1.41\times10^{-7}$ & $6.57\times10^{4}$  &  $2.96\times10^{-1}$ & 2.74$\times 10^{-3}$ & 4.45$\times 10^{-3}$\\
\hline 
\hline  
\end{tabular}

\caption{
The breakdown of Eq. (\ref{T_decomposition2}) at $E=5.5$ MeV for specific eigenstates, 
including the dominant eigen-mode (at $E_\lambda=5.4999$ MeV) shown in the bottom panel in Fig. \ref{Dattafull}. 
The table also lists the value of $\Gamma_\lambda$ defined by Eq. (\ref{gamma_perturbation}).  
\label{table:components}}
\end{table*}

\begin{figure}[htb]
\includegraphics[width=8.6cm]{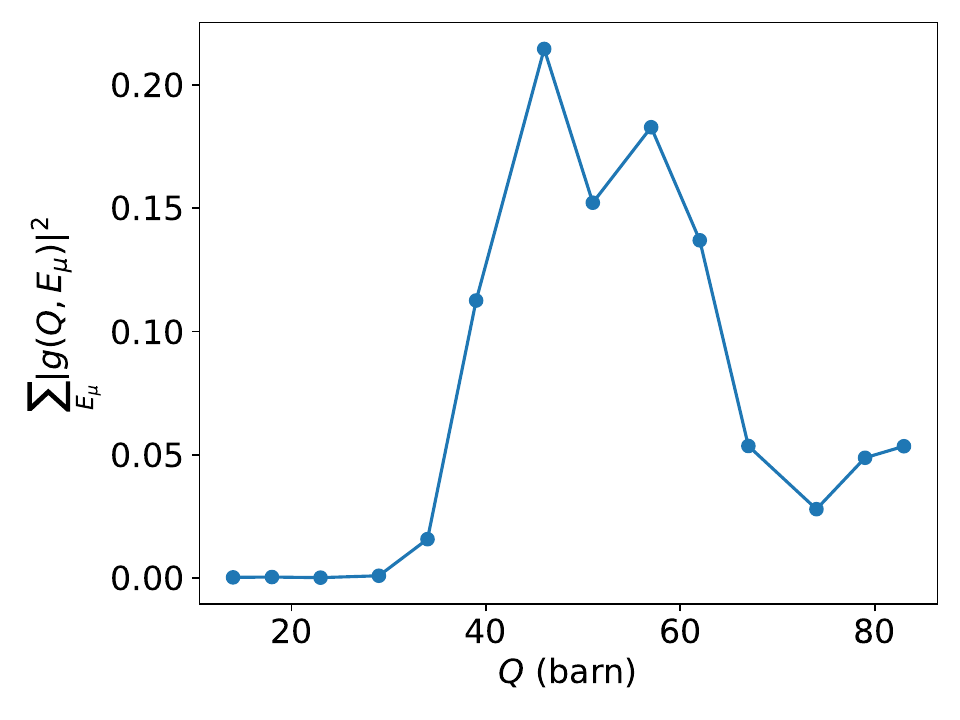}
\caption{
The square of the collective wave function $\sum_\mu |g(Q,E_\mu)|^2$ of the 
dominant eigen-mode for the transmission coefficient for fission. 
This is plotted as a function of $Q$, by adding all the excited configurations 
at each $Q$. 
}
\label{Datta6}
\end{figure}

The collective wave function for the dominant eigenstate is shown in Fig. \ref{Datta6}.
Here, the collective wave function is defined as,
\begin{equation}
g_j^{(\lambda)}=\sum_{j'}(N^{1/2})_{jj'}f_{j'}^{(\lambda)},
\end{equation}
where $N^{1/2}$ is the square root of the overlap kernel, $N$. 
In the figure, the square of the collective wave function is plotted as a function of $Q$, by summing all the configurations 
for each $Q$. 
One can see that this wave function has a peak in the middle of a chain of the $Q$-blocks, 
rather than at the position of the higher barrier, 
as would have been assumed in the transition state theory. 
This
can be easily understood if one uses a simple 3$\times$3 matrix with a tri-diagonal coupling, 
\begin{equation}
H=
\left(\begin{matrix}
     e_1 & v & 0 \cr
     v & e_2 & v' \cr
     0 & v' & e_3
           \end{matrix}\right). 
\end{equation}
When the off-diagonal couplings are zero, that is, $v=v'=0$, the three eigenvectors of this matrix read 
$\psi_1=(1,0,0)^T,~\psi_2=(0,1,0)^T$, and $\psi_3=(0,0,1)^T$. 
If the off-diagonal couplings are small, one can then use the first order perturbation 
theory. In this weak coupling limit, only the wave function $\psi_2$ acquires 
components both in the first and the third configurations. 
Therefore, the eigenstate which has siginificant components both in the first and the third 
configurations has the largest comonent in the second configuration. 
A similar argument can be applied when the dimension of the matrix is larger than 3. 

In this subsection, we have discussed 
the smallness of degrees of freedom, $\nu$, 
in terms of the transmission coefficient. 
See Appendix \ref{appendixRank}  and Ref. \cite{FUSION} for 
an alternative explanation of the smallness of $\nu$ based on the rank of 
$\Gamma_{\rm eff}$. 

\section{Summary and future perspectives \label{summary}}

We presented a novel approach to low-energy induced fission based on 
the method of non-equilibrium Green's function (NEGF), which has been widely used in problems of electron 
transport in condensed matter physics. To this end, we considered a model which consists of many-body 
configurations constructed with the constrained density functional theory. Compound 
nucleus configurations as well as pre-fission configurations were represented by random matrices.  
Transmission coefficients were then evaluated with the Datta formula in the NEGF formalism.  
We applied this method to neutron induced fission of $^{235}$U by restricting to seniority-zero 
configurations. We found that the fission-to-capture branching ratio was somewhat underestimated, even 
though the calculated value was still reasonable. As we chose the parameters as realistic as possible, this 
clearly indicated a necessity of seniority non-zero configurations. 
We also evaluated the number of degrees of freedom $\nu$ for fission. 
Our calculation yielded much smaller values for $\nu$ as compared to the number of 
the fission decay channels, which is 
consistent with the experimental data as well as the picture of transition state theory. 

We have argued that 
the smallness of $\nu$ can be explained 
in terms of the number of GCM eigenstates which 
significantly contribute to the transmission coefficient. 
While the smallness of $\nu$ has been explained based on the picture of the transition state theory, 
in this way 
the smallness of $\nu$ could be explained in a natural manner 
without assuming a priori the existence of transition states. 

We have found that there are three conditions for a GCM eigenstate to contribute significantly 
to transmission coefficients. 
Firstly, an eigenstate needs to have a large enough amplitude at the left-end configurations at $Q=Q_L$, 
at which the neutron width is defined. 
Secondly, it also needs to have a large enough amplitude at the right-end configurations at $Q=Q_R$, at which 
the fission width is defined. Lastly, the eigenenergy 
$E_\lambda$ has to be close to the excitation energy $E$ due to the Breit-Wigner factor in the 
transmission coefficient. 
While, in the transition state theory, transition states are assumed to locate at the barrier position, 
GCM eigenstates which satisfy all of these three conditions do not necessarily 
have the dominant component at the barrier position. 
In fact, in our calculation with a double humped barrier, we have found that the dominant 
eigen-mode has the largest component in between the two barriers. 

The method presented in this paper provides a promising way to microscopically understand 
nuclear fission. 
A big challenge is how to manage the dimension of 
Hamiltonian matrix, which increases rapidly as the model space increases.
In this regard, as we argued in this paper, one 
only needs a limited number of GCM eigenstates in order to 
compute transmission coefficients. One could then employ an 
iterative method, such as the Lanczos algorithm, 
to find a few eigenstates. With such a numerical technique, 
one could expand relatively easily 
the model space such that finite seniority 
configurations are also included. 
We will report on this in a separate publication \cite{Lanczos}.

As another future work, 
one can use the same model as the one presented in this paper to calculate 
a decay width for spontaneous fission and cluster decays \cite{hagino2020,hagino2020-2,uzawa2022}. 
It would be interesting to analyze how these decay modes 
are decomposed 
into eigenstates 
of the Hill-Wheeler equation.

\section*{Acknowledgments}

We thank G.F. Bertch for collaborations at the 
early stage of this work. 
We also thank D.A. Brown
for accessing the evaluated data in Ref. \cite{Brown20181short}.
This work was supported in part by JSPS KAKENHI
Grants No. JP19K03861 and JP23K03414, and JP23KJ1212. 
The numerical calculations were performed
with the computer facility
at the Yukawa Institute for Theoretical
Physics, Kyoto University.

\appendix

\section{Gaussian orthogonal ensemble and a chi-squared distribution\label{GOE}}

Here we show that pertubative decay widths in the
GOE follow  the chi-squared distribution for $\nu$ degrees of freedom
if the width matrix $\Gamma$ has equal non-zero eigenvalues and rank $\nu$, that is, 
\begin{equation}
\Gamma  = \left(\begin{matrix}
     \gamma&  &  & &  \cr
     & \ddots & &  &  \cr
      & &  \gamma&  &   \cr
      &  &  & 0&   \cr
      &  & & & \ddots \cr
           \end{matrix}\right).
\label{Gamma}           
\end{equation}
The proof is very simple.  In the GOE,  the
amplitudes $c_i = \langle n|i\rangle$ of the basis states $|i\rangle$ in
the eigenstates $|n\rangle$ follow the Gaussian distribution in the limit of a large matrix size\cite{RMP1981}.
Notice that, in the first order perturbation theory, 
the eigenenergy of the eigenstate $|n\rangle$ has an imaginary part of $-i
\langle n | \Gamma |n\rangle/2$, where 
the decay width is given by 
\begin{equation}
\langle n | \Gamma |n\rangle =\gamma \sum^\nu_{i=1} |c_i|^2. 
\label{gammaci}
\end{equation}
This quantity is given as a summation of the squares of Gaussian-distributed variables.
By definition, its distribution is the chi-squared distribution Eq. (\ref{P_nu}) with
$\nu$ degrees of freedom $\nu$.

\section{Rank of the matrix $\Gamma_{\rm eff}$\label{appendixRank}}

\begin{figure}
\includegraphics[width=8.6cm]{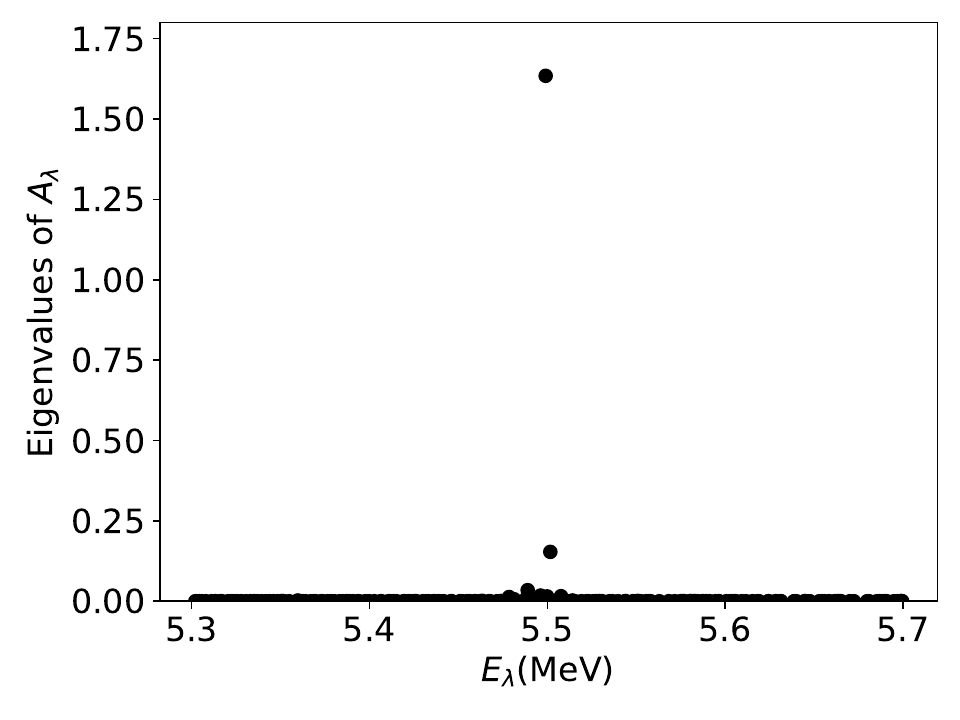}
\caption{
Eigenvalues of $A_n$ defined in Eq. (\ref{A_lambda}) as a function of eigenenergy $E_n$.
}
\label{Heff2}
\end{figure}

In Sec. III-C, we explained the small number of $\nu$ in terms of the spectrum decomposition of the Green function.
On the other hand, as we showed in Fig. \ref{distofEV2}, 
the matrix $\Gamma_{\rm eff}$ has a low-rank structure. 
In this Appendix we analytically evaluate the rank of $\Gamma_{\rm eff}$ to explain the smallness of 
$\nu$.

From Eq (\ref{eHamiltonian}), 
$\Gamma_{\rm eff}$ is given by, 
\begin{align}
(\Gamma_{\rm eff})_{i,j}&=2\sum_{kl}V_{ik}{\rm Im}((G_Q)_{kl})(V^T)_{lj},  
\end{align}
where $G_Q$ denotes the Green function corresponding to $H_Q$,
\begin{equation}
G_Q=(EN_Q-H_Q)^{-1}.    
\end{equation}
For simplicity of notation, we have used $V$ for $V^{(L)}$. 

Here we take the same procedure as in Sec. III-C and express $G_Q$ as
\begin{align}
(G_Q)_{kl}=\sum_\lambda O_{k \lambda}(\tilde{G}_Q)_{\lambda}O^T_{\lambda l}.
\end{align}
Here, $(\tilde{G}_Q)_{\lambda}$ denotes the $\lambda$-th eigenvalue of $G_Q$ and 
$O$ is defined as $O=\left(\vec{f}_1, \vec{f}_2, ... ,\vec{f}_N \right)$ with the column-vectors $\vec{f}_\lambda$ 
representing the GCM weight functions in Eq. (\ref{GCM}).
Then $\Gamma_{\rm eff}$ is transformed to 
\begin{align}
(\Gamma_{\rm eff})_{ij}&=\sum _{\lambda,E_\mu,E_{\mu'}} V_{i,(Q_1,E_\mu)} f_\lambda(Q_1,E_\mu)  \notag\\
&\times\frac{\Gamma_\lambda}{(E-E_\lambda)^2+(\frac{\Gamma_\lambda}{2})^2} \,f_\lambda(Q_1,E_{\mu'}) V_{j,(Q_1,E_{\mu'})},
\end{align}
where $Q_1$ is the first $Q$-block at $Q=18$ b, and $E_\mu$ denotes the label for the configurations at $Q_1$. 
Since the rank of a matrix $VAV^T$ is equal to the rank of the symmetric matrix $A$ \cite{Miller1990}, 
the rank of the matrix $\Gamma_{\rm eff}$  
is equal to the rank of  $\sum_\lambda A_\lambda$, defined as
\begin{equation}
(A_\lambda)_{\mu,\mu'}=f_\lambda(Q_1,E_\mu) \frac{\Gamma_\lambda}{(E-E_\lambda)^2+(\frac{\Gamma_\lambda}{2})^2}\,f_\lambda(Q_1,E_{\mu'}).
\label{A_lambda}
\end{equation}

Using the relation 
\begin{equation}
{\rm rank}\left(\sum_\lambda A_\lambda\right)\le \sum_\lambda {\rm rank}(A_\lambda),
\end{equation}
one can analyze the rank of each $A_\lambda$ separately.
Since the matrix $A_\lambda$ in Eq. (\ref{A_lambda}) has a separable form, 
it is a rank one matrix. 
This means that 
$A_\lambda$ has only one non-zero eigenvalue $a_\lambda$, which is equal to ${\rm Tr}(A_\lambda)$. 
$a_\lambda$ is evaluated as
\begin{align}
a_\lambda
&=\sum_{k} |f_\lambda(Q_1,E_k)|^2\frac{\Gamma_\lambda}{(E-E_\lambda)^2+(\frac{\Gamma_\lambda}{2})^2}\notag\\
&\simeq\left(\sum_{k} |f_\lambda(Q_1,E_k)|^2\right)\frac{\gamma_{{\rm fis}}(\sum_{E_l} |f_n(Q_{R},E_l)|^2)}{(E-E_\lambda)^2+(\frac{\Gamma_\lambda}{2})^2}.     
\end{align}
At the last line we have evaluated $\Gamma_\lambda$ with perturbation, see Eq. (\ref{gamma_perturbation}). 
This expression implies that only those eigenstates which have large enough weight at both $Q=Q_1$ and $Q=Q_R$ 
and whose eigenvalue $E_\lambda$ is close to the excitation energy $E$ contribute significantly 
to the rank of $\Gamma_{\rm eff}$.

The eigenvalues of $A_\lambda$, that is, 
$a_\lambda$ for our model at $E=5.5$ MeV 
are shown as a function of $E_\lambda$ 
in Fig. \ref{Heff2}.
One can see that most of $a_\lambda$ are almost zero, and only two of them have significant values. 
That is, only two matrices of $A_\lambda$ have rank 1 while the rest may be regarded to have rank 0. 
Therefore effectively $\sum_\lambda {\rm rank}(A_\lambda)$ is 2, which provides 
the upper limit of ${\rm rank}(\Gamma_{\rm eff})$ as ${\rm rank}(\Gamma_{\rm eff}) \le 2$.
This is a direct proof why the ${\rm rank}(\Gamma_{\rm eff})$ is small as shown in Fig. \ref{nu1}. 

\section{Distribution of eigenvalues of $\Gamma_{\rm eff}$ \label{App:eigenvalues}}

\begin{figure}
\includegraphics[width=8.6cm]{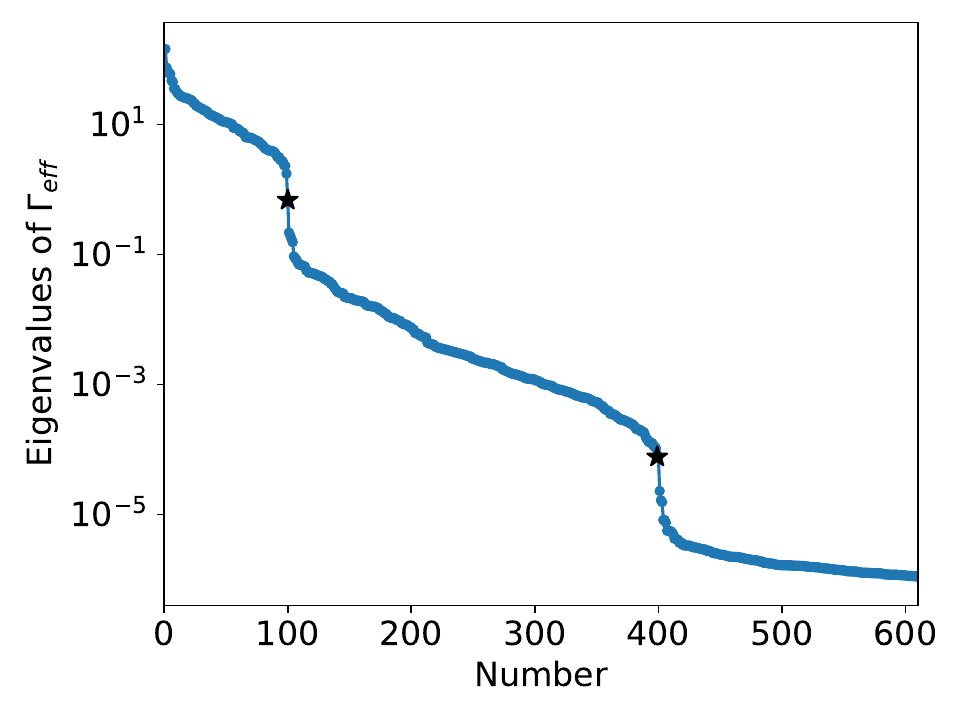}
\caption{Eigenvalues of $\Gamma_{\rm eff}$ at $E=6.5$ MeV for 100 different ensembles in descending order.
The 100th and 400th points 
are plotted by stars. 
}
\label{distofEV}
\end{figure}

In Fig. \ref{distofEV2}, we plotted the distribution of the eigenvalues of $\Gamma_{\rm eff}$ 
for a typical sample. 
We have generated 100 samples and confirmed that the feature of the distribution remains 
the same for the different random seeds. 
Fig. \ref{distofEV} shows the distribution of all those  
$10^3\times10^2=10^5$ eigenvalues in descending order. 
Reflecting the fact that there is one large and three intermediate eigenvalues in Fig. \ref{distofEV2}, 
the first 100 points and the subsequent 300 points form clusters.
The 100th and 400th points are marked with 
stars in the figure. 

\bibliography{nu}

%\end{thebibliography}
\end{document}